# FROM ENGLISH TO ASIC: HARDWARE IMPLEMENTATION WITH LARGE LANGUAGE MODEL *

**Emil Goh**
Engineering Product Development
Singapore University of Technology and Design
Singapore

**Maoyang Xiang**
Engineering Product Development
Singapore University of Technology and Design
Singapore

**I-Chyn Wey**
Department of Electrical Engineering
Chang Gung University
Taoyuan, Taiwan

**Tee Hui Teo**
Engineering Product Development
Singapore University of Technology and Design
Singapore
tthui@sutd.edu.sg

March 11, 2024

## ABSTRACT

In the realm of Application Specific Integrated Circuit (ASIC) engineering, the landscape has been significantly reshaped by the rapid development of Large Language Model (LLM), paralleled by an increase in the complexity of modern digital circuits. This complexity has escalated the requirements for Hardware Descriptive Language (HDL) coding, necessitating a higher degree of precision and sophistication. However, challenges have been faced due to the less-than-optimal performance of modern language models in generating hardware description code, a situation further exacerbated by the scarcity of the corresponding high-quality code datasets. These challenges have highlighted the gap between the potential of LLMs to revolutionize digital circuit design and their current capabilities in accurately interpreting and implementing hardware specifications.

To address these challenges, a strategy focusing on the fine-tuning of the leading-edge nature language model and the reshuffling of the HDL code dataset has been developed. The fine-tuning aims to enhance models' proficiency in generating precise and efficient ASIC design, while the dataset reshuffling is intended to broaden the scope and improve the quality of training material. The model demonstrated significant improvements compared to the base model, with approximately 10 to 20% increase in accuracy across a wide range of temperature for the pass@1 metric. This approach is expected to facilitate a simplified and more efficient LLM-assisted framework for complex circuit design, leveraging their capabilities to meet the sophisticated demands of HDL coding and thus streamlining the ASIC development process.

## 1 Introduction

The complexity associated with Application-Specific Integrated Circuits (ASIC) has seen a significant upsurge, demanding a more refined and sophisticated approach to Hardware Description Language (HDL). Modern processors are designed to perform highly specialized functions, and their development requires integrating billions of transistors into a single chip. For instance, the design and fabrication of a cutting-edge chip for a smartphone requires a detailed

---

orchestration of various components such as the CPU, GPU, memory controllers, and connectivity modules (Wi-Fi, Bluetooth, cellular networks) on a microscopic scale. This scenario exemplifies the heightened complexity in ASIC design, demanding a higher degree of precision and sophistication in every aspect of the chip's development, from conceptualization through to physical implementation.

In parallel, an Artificial Intelligence (AI) research company called OpenAI shook the world by introducing ChatGPT in Year 2022, a powerful generative AI capable of producing natural language based on text prompts. As the technology has advanced over the past year, generative AI is now capable of generating not only natural language, but also images, sounds, and even code. Since the release of ChatGPT, competition in the field has been intense and rapidly evolving. Major companies like Google, Meta, and Amazon have invested heavily in generative AI technology, with most developing their own large language model (LLM) and integrating the capabilities into their products and services. Such advancements have not only enhanced the capabilities of natural language processing but also underscored their potential to transform a variety of sectors. This convergence of advancements presents a promising field for the application of LLM in ASIC design, offering a novel approach to tackling the intricate challenges of HDL coding.

Nevertheless, the implementation of LLM in HDL coding has confronted significant obstacles, particularly evident in their efficacy in generating precise coding outputs. Beyond the inherent challenges posed by the intricacies of circuit design and the high precision necessitated therein, generative AI models are prone to producing syntax errors within Verilog codes. Such discrepancies become increasingly pronounced in scenarios involving complex design architectures or specific, hardware-oriented syntax. While these errors might appear minor, they possess the potential to engender incorrect hardware behavior or provoke complications during the simulation and compilation phases. This issue is exacerbated by the scarcity of high-quality code datasets and exemplars, as demonstrated by the limited availability of intricate Verilog code examples. These examples are indispensable for the effective training of LLMs, enabling them to accurately comprehend and execute the nuanced requirements essential for ASIC design.

To tackle these issues, a two-fold approach has been proposed: fine-tuning of forefront LLMs to improve their efficacy in HDL coding, coupled with the expansion of the coding dataset to encompass a more diverse array of coding examples and complexities. This paper endeavors to explore the application of generative AI in the production of HDL, with a particular focus on Verilog. In pursuit of this objective, the Mistral 7B Large Language Model (LLM) will undergo a process of fine-tuning, aimed at augmenting its proficiency in generating Verilog code with heightened effectiveness[1].

The organization of this paper follows a logical progression, starting with an exploration of the current developments in LLMs in Section 2. The following section details our methodology for fine-tuning the model specific to HDL. Section 4 evaluates the effectiveness of the proposed model, which is followed by summarizing the advantages and limitations of our approach in Section 5. This structure aims to comprehensively present our innovative approach towards leveraging LLM technology for a more efficient and simplified process in ASIC design.

## 2 Literature Review

Recent advancements in the field of AI, specifically in LLMs have shown remarkable capabilities in understanding and generating human-like text. Among these, Mistral 7B, a 7-billion-parameter LLM, presents a promising avenue for application in various domains, including the generation of HDL code, which is crucial for modern chip design[1]. This section review explores the background, LLM, and the potential of fine-tuning Mistral 7B to generate Verilog code, a popular HDL used in digital circuit design.

### 2.1 Generative AI and LLMs

Generative AI encompasses a broad spectrum of AI algorithms designed for the creation of novel content through the processing of extensive training datasets. Within this domain, LLMs stand out as a pivotal subset, providing a remarkable capability for generating text that is both coherent and contextually relevant. Its transformative impact on the field was notably augmented by the introduction of the transformer architecture and the self-attention mechanism by Google in 2017[2]. The transformer, a neural network architecture that relies exclusively on self-attention, has significantly advanced language understanding capabilities. It has demonstrated superior performance over traditional recurrent and convolutional models across various translation benchmarks. Its encoder-decoder configuration architecture maps input sequences to continuous representations, which are then processed by the decoder to generate output sequences in an auto-regressive fashion. This means each subsequent element is produced based on the preceding elements, leveraging layers of self-attention and fully connected layers for efficient operation.

The advent of transformer architecture has catalyzed the development of Generative AI, marking a departure from the predictable outcomes of conventional deterministic AI algorithms. As a form of stochastic AI, Generative AI thrives on the diversity of training data spanning multiple domains, enabling the generation of inventive content across text, images, audio, and synthetic data realms that align with the input prompt's context.

Table 1: Mistral 7B model architecture

| Parameter | Value |
|---|---|
| dimension | 4096 |
| model_layers | 32 |
| attention_head_dimension | 128 |
| hidden_dimension | 14336 |
| number_of_heads | 32 |
| number_of_key-value_heads | 8 |
| window_size | 4096 |
| context_length | 8192 |
| vocabulary_size | 32000 |

An example of innovation within this space is the Mistral 7B model, developed by Mistral AI. This 7-billion-parameter, open-source language model has been designed to balance high-level performance with efficient inference, claiming to outperform larger models from notable entities like Meta in various benchmarks. Notably, it has excelled in generating functional code and solving mathematical tasks, surpassing the LLaMa 34B model. Mistral 7B leverages advanced attention mechanisms, such as grouped-query attention (GQA) and sliding window attention (SWA), to enhance inference speed and reduce memory requirements for decoding, effectively handling long sequences with lower computational demand[3, 4, 5]. The Mistral 7B model builds on the transformer architecture. The model's parameters are summarised in Table 1.

The selection of the Mistral 7B model for fine-tuning in this research, aimed at generating Verilog code, underscores its efficacy and applicability for supporting code generation tasks. Its open-source nature and exceptional performance make it an ideal candidate for the objectives of this study.

## 2.2 Fine Tuning of Large Language Model

Fine-tuning involves additional training on a pre-trained model with a smaller dataset to perform a specific task. This section highlights the different concepts and methods that are essential for the fine-tuning of LLM.

### 2.2.1 Transfer Learning

Transfer learning is the crucial technique for pre-trained models to be re-purposed on another related task with minimal additional training through the transfer of knowledge [6]. This technique enables existing trained models to be used as a foundation for another task's development.

Using the method of transfer learning, enormous resources that are required to train LLM could be saved. However, transfer learning only works if the initial training task is general [7]. This form of learning mechanism by improving performance through the learning of a different but related knowledge is also known as inductive transfer learning.

### 2.2.2 Low-Rank Adaptation (LoRA)

LLMs are known to contain billions of parameters and it is not efficient and effective to train LLMs from scratch during fine-tuning. Low-Rank Adaptation or LoRA [8] is a technique used to accelerate the fine-tuning of pre-trained models with minimal computational overhead and memory consumption.

The LoRA approach freezes the original pre-trained weights, identifies adaptation points, and introduces low-rank matrices into the different layers of the model's architecture. These matrices can later be trained with the new data while keeping the original weights frozen. The original and adapted weights are then merged to produce the fine-tuned model. The LoRA approach is the common method of fine-tuning and is used in the Parameter Efficient Fine-Tuning (PEFT) library from Hugging Face.

### 2.2.3 DeepSpeed ZeRO

DeepSpeed is a library designed to optimize training processes and facilitate the fitting of large models onto GPUs[9]. It is powered by Zero Redundancy Optimiser (ZeRO), which is an algorithm used to enhance communication between

Figure 1: Preview of the processed dataset on HuggingFace

GPUs. It also addresses common challenges faced by LLM fine-tuning such as memory constraints and slow training times.

The work presented in this paper utilizes DeepSpeed ZeRO Stage 3 offloading capabilities [10]. This enables the training of very large models on limited GPU resources by leveraging on the GPU, CPU, and NVMe to enable scalability.

### 2.2.4 Quantization

Quantization techniques are used commonly in the training and inference of LLMs. These methods reduce memory usage and computational cost by representing parameters at lower precision. Utilizing quantized data types allows users to load larger models into limited memory space and speed up model inference. As LLMs are massive in size, half-precision floating point format (FP16) is commonly used for training and inference. It uses 16 bits instead of the usual 32 bits (FP32) for single precision.

In this section, various mechanisms that enable the LLMs and generative AI were introduced. On top of that, methods to optimize fine-tuning processes of LLMs were discussed which lay a solid foundation for the proposed language model for LLM.

## 3 Method

Moving on to the LLM and generative AI portion of the implementation flow, this section looks into the fine-tuning of existing LLM to generate HDL, specifically Verilog.

### 3.1 Dataset

The lack of labels in existing datasets causes less quality LLM to be produced and hinders many future developments. This includes the utilization of fine-tuning methods such as Direct Preference Optimisation (DPO). To solve this issue, OpenAI's GPT-3.5 Turbo was employed to label the unlabelled datasets found on Hugging Face's repositories.

To begin, a Hugging Face dataset repository was chosen and processed before deploying the GPT-3.5 Turbo API to label the data entries. In this case, a Verilog dataset, wangxinze/Verilog_data [11], was selected for processing as the dataset consists of a single module per entry but is not well-labelled.

The dataset, in CSV format, was first converted into a data frame using the Pandas Python data analysis library for easier manipulation. Using Python's regular expression operation, the modules within each entry are extracted such that each entry would start with 'module' and end with 'endmodule'. Duplicates and empty entries are removed to reduce biases when training LLM. Next, each module's name was extracted and placed in a separate column, corresponding to the rows, for annotation. The GPT-3.5 Turbo API was finally deployed after this step to generate a short description for each of the data entries.

The result of this process is a Verilog dataset with 68,122 data entries that are written by humans. This is by far the largest labeled Verilog dataset available on Hugging Face and probably online. A preview of the dataset is shown in Figure 1.

Table 2: Parameters used for Fine-Tuning

| Parameter | Value |
|---|---|
| learning_rate | $1e-5$ |
| adam, $\beta_1$ | 0.9 |
| adam, $\beta_2$ | 0.999 |
| adam, $\epsilon$ | $1e-8$ |
| number of training epochs | 1 |
| gradient checkpoint | True |
| training batch size per device | 1 |
| bf16 | True |
| lr scheduler type | cosine |
| warmup ratio | 0.1 |
| max sequence length | 2048 |
| rank | 64 |
| lora alpha | 16 |
| bias | none |
| task type | CAUSAL_LM |
| zero optimization stage | 3 |
| offload optimizer device | CPU |
| offload param device | CPU |

The processed dataset can be found here: https://huggingface.co/datasets/emilgoh/verilog-dataset-v2

### 3.2 Training Process

Mistral 7B is selected for fine-tuning as it is open-source and capable of supporting code[1]. The parameters used for the fine-tuning can be found in Table 2. To optimize the training process, some measures and strategies were taken.

#### 3.2.1 Checkpoint and Evaluation

Unfortunately, there is no good solution to evaluate code generated automatically at this point. Evaluating the generated code would involve the calling of an external simulator to run the code and check for errors. This is complicated to implement when training is going on at the same time. Hence, training checkpoints are used such that the model can be evaluated from time to time at different training steps.

For this work, checkpoints were saved and evaluated at steps 10,000, 20,000, 40,000, and 60,000. At these checkpoints, inferences were carried out and 10 samples were generated with a temperature of 0.2 for each problem in the VerilogEval-Machine benchmark descriptions. Verilog modules in the generated samples were roughly extracted and evaluated using the VerilogEval-Machine benchmark. Although the data were not carefully cleaned, it gives a rough gauge of the model's accuracy at the particular step and warns of an unlikely case of overfitting.

Apart from evaluations, implementing training checkpoints also prevents training progress from being lost due to runtime errors. With this measure in place, users can retrieve trained data from the last saved checkpoint and continue with the training process instead of starting over.

#### 3.2.2 DeepSpeed ZeRO

Fine-tuning a large language model requires large RAM space. DeepSpeed ZeRO [9] was implemented during training to allow the checkpoint weights to be offloaded to the CPU which has more RAM space but lower latency. The offloaded weights are then moved from the CPU to the GPU, which has less RAM space but high latency, as the input reaches the different layers. This helps optimise memory utilization in the system and prevent runtime errors due to the lack of memory space to occur.

### 3.3 Inference Strategy

The fine-tuned model is later deployed on Google Colaboratory's A100 GPU again and used for inference to generate samples for evaluation. Just like training, model inference takes time and requires high VRAM. Hence, Google

Colaboratory was used to gain access to NVIDIA's A100 GPU. The following are some strategies adopted to accelerate the speed of inference and optimize memory usage including caching, flash attention, and token limits.

#### 3.3.1 Caching

Caching during inference enables the model to save and reuse intermediate hidden states from previous time steps. As the generation of new tokens depends on what was generated previously, caching enables the model to avoid recomputation and speeds up the inference process. By enabling cache, memory usage for inference could also be optimized. Large, duplicated intermediate states do not have to be stored again and could benefit when generating long sequences.

#### 3.3.2 Flash Attention

Flash attention [12] is enabled to reduce the time taken to generate the samples. It is used to optimize transformer-based models specifically by improving the operations of the attention mechanism.

High bandwidth memory (HBM) has high memory capacity, but is slow in processing. On the other hand, SRAM has a fast processing speed, but smaller memory capacity. Hence, the standard attention mechanism often uses HBM to store, read, and write keys, queries, and values, and the GPU on-chip SRAM to perform the attention mechanism. The process involves the loading of the elements from the HBM to the SRAM, and writing the results back to the HBM, and is repeated for every attention step.

Flash attention optimizes the operation by loading the aforementioned elements only once, reducing the writing and loading process. It fuses the operations of the attention mechanism and finally writes the results back after the completion of the fused operation.

#### 3.3.3 Token limits

Lastly, the token limit of the generated samples is limited to 500 tokens. From early experiments, it was found that the LLM would sometimes generate the corresponding testbench which is not needed for evaluation and could result in difficult extraction of the targeted Verilog code. By limiting the number of generated tokens, the excessive generation of unwanted tokens after the sample Verilog code is prevented. This also speeds up the generation of the samples for evaluation later.

### 3.4 Post-Processing

After the generation of samples, regular expression operations in Python are used to extract the Verilog module. The operation searches the sample for the keywords, 'module' and 'endmodule', before extracting the text including the two keywords. Next, the keywords, 'module' and ';', are searched for and removed from the extracted text. The final output will be the module's logic ending with 'endmodule'. This is the format required for VerilogEval, which is the benchmark used for evaluation.

This section highlights the processes and measures taken in the labeling of the dataset containing Verilog code using GPT-3.5 API, the fine-tuning of Mistral 7B to generate Verilog code, and the inference process of the model to generate samples for further evaluation.

## 4  Result

This section evaluates the fine-tuned model's ability to generate Verilog code. It begins with an overview of the VerilogEval benchmark, employed for this assessment[13]. The model's performance is scrutinized at various stages and its final output through the VerilogEval-Machine benchmark[13]. A comparison with similar models in the field provides a comprehensive analysis. Lastly, an examination of selected code samples from both the proposed and foundational models identifies errors, offering insights for future improvements.

### 4.1  VerilogEval

VerilogEval [13] is a benchmark for LLMs to evaluate the capability of Verilog code generation by a research team from NVIDIA. It contains two parts to the evaluation set - Machine and Human. Both evaluation sets contain problems from HDLBits [14] which is a website with digital circuit design exercises to learn Verilog HDL.

As questions on the HDLBits website are in the form of images, diagrams, and tables, they are not compatible with non-multimodal LLMs. Thus, gpt-3.5-turbo was employed by the research team to automate the generation of the problem descriptions according to each problem's answer. This forms the description-answer pairing for VerilogEval-Machine. The final output was a total of 143 problems available for evaluation.

Similarly, VerilogEval-Human consists of problems from HDLBits like VerilogEval-Machine. However, in this case, the researchers manually examined the problem descriptions on the website and converted them into text. VerilogEval-Human contains a total of 156 problems, 13 more problems than VerilogEval-Machine as the conversion was done manually.

As mentioned before, VerilogEval uses the same evaluation metric as HumanEval but with fewer generated samples per problem for evaluation. In VerilogEval, the number of generated samples for evaluation is 20, while HumanEval uses 200 generated samples for evaluation. Apart from that, the VerilogEval benchmark requires the use of iVerilog for the simulation of the generated samples.

### 4.2 Evaluation with VerilogEval-Machine

VerilogEval[13] consists of two sections: Machine and Human. For the purpose of this study, VerilogEval-Machine is used for evaluation. VerilogEval-Machine is particularly valuable in assessing a model's capability in comprehending instructions and generating Verilog code that is functional and syntactically correct. Also, the dataset used for training does not contain text or code that is produced extensively by GPT-3.5, thus minimizing the chances of having positive results due to similarities in questions and answers within VerilogEval-Machine's problem sets.

#### 4.2.1 Evaluation at Checkpoints

The model is evaluated at different checkpoints to observe its performance against training loss. However, the evaluation of generated samples is just a gauge and not accurate. The evaluation results of each checkpoint step can be seen in Table 3, where the metrics *pass@k* is used to evaluate the model's performance, specifically in tasks such as code generation or answering questions which is introduced by HumanEval[15]. These metrics measure the model's ability to produce correct or satisfactory answers within a specified number of attempts or samples. It uses the *pass@k* metric to evaluate functional correctness. Equation 1 is used to represent the evaluation metric.

$$pass@k := \sum_{problems} \left[1 - \frac{\binom{n-c}{k}}{\binom{n}{k}}\right] \quad (1)$$

Where:

- $n$: the number of samples generated per task
- $c$: the number of samples that pass test, $c \leq n$
- $k$: evaluation budget

Using the *pass@k* metric, a problem is marked as solved if at least one of the generated samples passes the test within the evaluation budget.

Table 3: Evaluation of the Model at Different Checkpoints

|                  | VerilogEval-Machine |         |          |
|------------------|---------|---------|----------|
| **Checkpoint Steps** | **pass@1** | **pass@5** | **pass@10** |
| 10,000           | 10.84 % | 20.67 % | 25.87 %  |
| 20,000           | 10.07 % | 21.94 % | 25.87 %  |
| 40,000           | 3.43 %  | 9.12 %  | 11.89 %  |
| 60,000           | 11.26 % | 23.39 % | 28.67 %  |

The results produced at step 40,000 observed a deterioration in evaluation results, compared to the previous checkpoint steps. This phenomenon could be a sign of overfitting, where loss continues to decrease while accuracy does not improve. However, the results are just a gauge and could be inaccurate. Nonetheless, training continued and an increment of accuracy was observed at 60,000 checkpoint steps.

arXiv TemplateA PREPRINT

arXiv Template  A PREPRINT### 4.2.2 Evaluation of Final Results

For each of the 143 problems in VerilogEval-Machine, 20 samples are generated and samples are generated at temperature = {0.2, 0.5, and 0.8}. This results in a total of 2860 samples generated at each temperature setting. This process is done for both the proposed and the base model.

Table 4 shows the accuracy of both models at various temperatures across different $k$ values. From the results, the proposed model shows significant improvements, outperforming the base model across the different temperatures. The improvements in accuracy is also evident across different $k$ values. Table 5 presents a concise overview of the improvements achieved by the proposed model, in comparison to the base model.

Table 4: Summary Results from VerilogEval-Machine

| Model | Temperature | VerilogEval-Machine | | |
|---|---|---|---|---|
| | | $k=1$ | $k=5$ | $k=10$ |
| **Proposed Model** | 0.2 | 40.59% | 43.42% | 44.24% |
| | 0.5 | 28.29% | 48.35% | 54.39% |
| | 0.8 | 19.93% | 42.44% | 50.64% |
| **Mistral-7B-v0.1** | 0.2 | 18.11% | 34.86% | 41.19% |
| | 0.5 | 10.77% | 34.06% | 46.02% |
| | 0.8 | 4.58% | 18.94% | 30.68% |

Table 5: Summary of Improvement in Proposed Model Compared to Mistral 7B

| Model | Temperature | VerilogEval-Machine | | |
|---|---|---|---|---|
| | | $k=1$ | $k=5$ | $k=10$ |
| **Proposed Model** | 0.2 | + 22.48% | + 8.56% | + 3.05% |
| | 0.5 | + 17.52% | + 14.29% | + 8.37% |
| | 0.8 | + 15.35% | + 23.50% | + 19.96% |

Evaluation of Results by Circuit Type The evaluation of the VerilogEval-Machine results is further extended into their respective circuit types. Table 6 outlines the number of circuit types questions in the VerilogEval-Machine problem set, the number of attempts made by the two different models, and the number of questions answered successfully by the models.

Next, Table 7 provides a summary of the models' accuracy for each circuit type in percentage. Although significant improvements are observed across the various circuit types, the accuracy for "shift register" related problems did not improve. Apart from that, problems with complex circuit types such as "finite state machines" ("fsm") and "larger circuits", which is a combination of different circuit types, saw little improvements in accuracy.

Comparison with Other Work During the course of this research, another research group published a work that also focused on the fine-tuning of Mistral 7B to generate Verilog code. The research introduced a new fine-tuned model, RTLCoder [16], which was trained on a synthetic dataset comprising questions and their corresponding answers, generated by GPT-3.5. The accuracy of the proposed model at different temperatures and the best result from the RTLCoder are detailed in Table 8. The summarised table reveals that the proposed model is approximately 20% behind the RTLCoder in terms of accuracy.

The discrepancy in accuracy could stem from the different inference methods. Apart from that, there could be overlaps in the GPT-3.5 generated questions in VerilogEval-Machine and the generated dataset used to train RTLCoder. Additionally, the process of generating the synthetic dataset, which contains both questions and Verilog code, is likely to have incurred more cost than the use of GPT-3.5 to label the dataset in this work's approach.

Nonetheless, RTLCoder presents an impressive achievement in using synthetic Verilog code to fine-tune existing LLM.

Table 6: Evaluation of Model by Circuit Type

| Circuit Type | Number | Tries | Proposed Model | Mistral-7B |
|---|---|---|---|---|
| arithmetic | 4 | 240 | 95 | 52 |
| basics | 27 | 1620 | 901 | 149 |
| cellular automaton | 2 | 120 | 50 | 6 |
| counters | 4 | 240 | 97 | 66 |
| debugging | 3 | 180 | 54 | 31 |
| features | 5 | 300 | 114 | 33 |
| fsm | 26 | 1560 | 61 | 41 |
| kmap | 8 | 480 | 126 | 51 |
| larger circuits | 7 | 420 | 17 | 15 |
| latches and flip-flops | 18 | 1080 | 224 | 120 |
| multiplexers | 5 | 300 | 143 | 38 |
| procedures | 8 | 480 | 156 | 72 |
| shift registers | 7 | 420 | 1 | 5 |
| vectors | 9 | 540 | 193 | 67 |
| waveform | 10 | 420 | 317 | 177 |

Table 7: Evaluation of Model by Circuit Type

| Circuit Type | Proposed Model | Mistral-7B | Difference |
|---|---|---|---|
| arithmetic | 39.58% | 21.67% | + 17.92% |
| basics | 55.6% | 9.20% | + 46.42% |
| cellular automaton | 41.67% | 5.00% | + 36.67% |
| counters | 40.42% | 27.50% | + 12.92% |
| debugging | 30.00% | 17.22% | + 12.78% |
| features | 38.00% | 11.00% | + 27.00% |
| fsm | 3.91% | 2.63% | + 1.28% |
| kmap | 26.25% | 10.63% | + 15.63% |
| larger circuits | 4.05% | 3.57% | + 0.48% |
| latches and flip-flops | 20.74% | 11.11% | + 9.63% |
| multiplexers | 47.67% | 12.67% | + 35.00% |
| procedures | 32.50% | 15.00% | + 17.50% |
| shift registers | 0.24% | 1.19% | - 0.95% |
| vectors | 35.74% | 12.41% | + 23.33% |
| waveform | 52.83% | 29.50% | + 23.33% |

Table 8: Evaluation of Model with Other Work

| | Temperature | pass@1 | pass@5 | pass@10 |
|---|---|---|---|---|
| Proposed Model | 0.2 | 40.59% | 43.42% | 44.24% |
| | 0.5 | 28.29% | 48.35% | 54.39% |
| | 0.8 | 19.93% | 42.44% | 50.64% |
| RTLCoder | Best | 62.5% | 72.2% | 76.6% |

## 4.3 Case study of generated samples

Erroneous responses from LLM are expected. Through different case studies, future work and improvements can be made to the dataset, training process, or inference process of the model. Different case studies also demonstrate that the final accuracy of the different models shown in the later part of this section might be less than the models' true accuracy.

### 4.3.1 Erroneous Response by Proposed Model

Case studies of erroneous responses generated by the proposed model are highlighted in this sub-section.

Among the samples generated by the proposed model, a common error encountered was the omission of the 'endmodule' keyword at the end of module declarations. In the example below, the generated Verilog code for a 'NOT' gate would have been considered correct if it had included the 'endmodule' keyword at the end. Apart from that, the lack of 'endmodule' keyword results in difficulties extracting the Verilog modules for evaluation during post-processing.

```
{   "task_id": "notgate",
    "completion":
        "// Verilog code:
            assign out = ~in;
        // End of file"
}
```

The fine-tuned model is also susceptible to hallucination and produces nonsensical outputs in some cases. Below is an example where the comment, 'Verilog code', is repeated redundantly multiple times without generating meaningful code. Similar issues were also observed in the generated samples of the base model and will be highlighted in the later sub-section.

```
{   "task_id": "dff8r",
    "completion":
        "// Verilog code
         // Verilog code"
}
```

Another common issue is the model attempting to instantiate undefined modules to solve particular problems. This phenomenon leads to a syntax error when the code undergoes simulation.

```
{   "task_id": "bugs_mux2",
    "completion":
        "   // Verilog code:

            // Instantiated modules
            inst_module inst_module (
                input sel,
                input [7:0] a,
                input [7:0] b,
                output reg [7:0] out
                );
            // Port connections
            assign inst_module.sel = sel;
            assign inst_module.a = a;
            assign inst_module.b = b;
            assign out = inst_module.out;
        // End of top module
```

```
18              endmodule",
19      "result": "failed: syntax error.",
20      "passed": false
21  }
```

#### 4.3.2 Erroneous Response by Mistral 7B

Next, common erroneous responses by the base model are presented.

One common type of erroneous response by the base model is syntax errors. As the base model has not been exclusively trained on Verilog codes, it could confuse the syntax rules of Verilog with those of other languages. Below is an example of a generated sample with syntax error by the base model. In this example, '#' is included in every line which caused an error.

```
1   {   "task_id": "mt2015_q4b",
2       "completion":
3           "   module(
4                   # input x,
5                   # input y,
6                   # output z);
7                   # always@(*)
8                       # begin
9                       # z = (x ^ y);
10                      # end
11              # endmodule",
12      "result": "failed: syntax error.",
13      "passed": false
14  }
```

In some cases, despite the clear instruction to generate a Verilog module, the generation of code in other programming languages was observed. The following illustrates an instance of such case. In this example, the base model produced a function in the C language instead of Verilog.

```
1   {   "task_id": "dff8ar",
2       "completion":
3           "   [SOL]
4               #include<stdio.h>
5               int main()
6               {
7                   int i,j,k,l,n;
8                   int a[100],b[100];
9                   printf('Enter the number of elements:');
10                  scanf('%d',&n);
11                  printf('\nEnter the elements of the first array:');
12                  for(i=0;i<n;i++)
13                  {
14                      scanf('%d',&a[i]);
15                  }
16                  for(i=0;i<n;i++)
17                  {
18                      for(j=i+1;j<n;j++)
19                      {
20                          if(a[i]>a[j])
```

```
21                   {
22                       k=a[i];
23                       a[i]=a[j];
24                       a[j]=k;
25                   }
26               }
27           }
28           printf('\nSorted array in ascending order:');
29           for(i=0;i<n;i++)
30           {
31               printf('%d',a[i]);
32           }
33           printf('\nEnter the number of elements:');
34           scanf('%d',&n);
35           printf('\nEnter the elements of the second array:');
36           for(i=0;i<n;i++)
37           {
38               scanf('%d',&b[i]);
39           }
40           for(i=0;i<n;i++)
41           {
42               for(j=i+1;j<n;j++)
43               {
44                   if(b[i]<b[j])
45                   {
46                       k=b[i];
47                       b[i]=b[j];
48                       b[j]=k;
49                   }
50               }
51           }
52           printf('\nSorted array in descending order:');
53           for(i=0;i<n;i++)
54           {
55               printf('%d',b[i]);
56           }
57           return 0;
58       }
59       [/SOL]"
60   }
```

Other than that, as mentioned in the previous sub-section, some of the generated samples exhibited instances of hallucination. Often, the instructions given were re-generated before coming to an abrupt stop. This phenomenon might be due to the token limit. However, it is important to note that the instances of hallucination are unlikely to be caused by the inference's token limit as the limit only accounts for new tokens generated.

```
1   {
2       "task_id": "mux256to1",
3       "completion": "Generate Verilog code for the following task:
4       This Verilog module is a top-level module that takes in an
5       input vector of 256 bits (in) and an 8-bit selection vector
6       (sel) and outputs a single bit (out). The output bit is
7       determined by the selection vector, which is used to select
8       one of the 256 bits from the input vector. The output bit is
```

```
 8          the bit from",
 9      "result": "failed: syntax error.",
10      "passed": false
11  }
```

Lastly, there were also cases where the base model only generated the skeleton of the code, without module logic. This can be seen in the following example.

```
 1  {   "task_id": "always_nolatches",
 2      "completion":
 3      "   module top_module (
 4              input [15:0] scancode,
 5              output reg left,
 6              output reg down,
 7              output reg right,
 8              output reg up
 9          );
10              // Your code goes here
11          endmodule"}
```

Through the analysis of case studies involving erroneous responses, valuable insights on improving the methods of preparing training datasets and training methods have been gained.

This section concludes with the evaluation of the fine-tuned model for Verilog code generation. The model demonstrated significant improvements compared to the base model, with approximately 10 to 20% increase in accuracy across a wide range of temperature for the pass@1 metric.

## 5  Conclusion

This work also produced possibly the first labeled Verilog dataset with code written by humans. This dataset is also likely to be the largest labeled Verilog dataset available on Hugging Face, containing a total of 68,122 data entries. Meanwhile, this research showcased a model that is fine-tuned from Mistral 7B to generate Verilog code. Through the evaluation, the model achieved approximately 15 to 20% improvement in accuracy for pass@1 at various temperatures compared to the base model.

Results have shown that there are still imperfections in the fine-tuned model and more can be done to improve the quality of Verilog code produced. Hallucinations still persist in the generated samples. A potential solution could be shortening the context length of the dataset. As such, the attention mechanism could better manage the relationship between the label and the different parts of the input. Shortening the context length would also lead to fewer entries in the dataset, thus allowing more training epochs to be done during fine-tuning.

An advantage of having a labelled training dataset is that data can be filtered more easily and fine-tuning efforts could be more targeted. In this study, the model was fine-tuned to address diverse circuit types and problem sets. By organizing and categorizing the dataset according to the circuits' functionality, the fine-tuned model could be specialized at handling specific circuit types and possibly yield more functional and valid results.

Apart from that, the proposed fine-tuned model in this work saw little to no improvements for problems that require more logical reasoning. This issue could be resolved by integrating the model with retrieval augmented generation (RAG) which is a database that LLMs could refer to for few-shot learning.

Lastly, although the fine-tuned model is published on Hugging Face, it is at the moment unable to be deployed locally and used in chat mode. Quantizing and converting the model into the GPT-Generate Unified Format (GGUF) would allow users to run the model locally and design circuits through conversation. This also enables more functionally correct and possibly complicated circuits to be generated as few-shot and chain-of-thought prompting could be utilized.

The proposed implementation flow in this work leverages the latest advancements in LLM and generative AI. While these technologies may already seem to be powerful, generative AI is still in its early phase of development and this is just the beginning of a new technological revolution. There are still many possibilities and much more work to be done in bridging NLP and IC design.

## Acknowledgments

We would like to thank SUTD-ZJU IDEA Visiting Professor Grant (SUTD-ZJU (VP) 202103, and SUTD-ZJU Thematic Research Grant (SUTD-ZJU (TR) 202204), for supporting this work.